\documentstyle[preprint,aps,12pt]{revtex}
\begin{document}

\title{Statistical Theory of 2-Dimensional Quantum Vortex Gas:\\
Non-Canonical Effect and Generalized Zeta Function}

\author{Hideki Ono and Hiroshi Kuratsuji}

\address{Department of Physics, Ritsumeikan University \break
Kusatsu City, 525-77, Japan}

\date{\today}
\maketitle

\begin{abstract}
The purpose of this paper is to present a quantum statistical
theory of 2-dimensional vortex gas based on
the generalized Hamiltonian dynamics recently developed.
The quantized spectrum is evaluated for a pair of vortex on
the basis of the semiclassical
quantization rule. This is used to evaluate the partition
function for a  dilute vortex gas. A remarkable consequence is that
the partition function and related quantities are given
in terms of the generalized Riemann zeta function.
The topological phase transition is naturally understood
as the pole structure of the zeta function.
\end{abstract}

\bigskip
\bigskip

\noindent
{\large\bf Introduction}

The study of vortex defects in two dimension has revealed many
interesting properties which have had a wide variety of applications
in condensed matter physics. For example the
superfluid ${}^4$He plays an important role in understanding
the properties of the physics of vortex defects. Apart
from the quantum fluid such as ${}^4$He,
it has long been known that in classical hydrodynamics
the equation of motion for the center of vortices forms
a Hamiltonian system, (e.g. Kirchhoff equation)
\cite{Lamb}
where the fluid under consideration is assumed to be
in an effectively two-dimensional incompressible fluid.
In our recent papers we have developed a novel theory of
vortex motion occurring in the two-dimensional quantum
condensates, namely, the superfluid liquid Helium
\cite{Kra}
as well as the Heisenberg spin model
\cite{Ono}.
This theory shows that the vortex motion is described by
the generalized Hamiltonian dynamics, which indicates the
deviation from the conventional canonical form.
Here we have used the time dependent Landau Ginzburg (LG)
equation (action) to derive the effective action for
the motion of center of vortices.
The LG action is described by the complex order parameter field
which is naturally introduced through the use of the
coherent state path integral for the quantum condensates.

On the other hand,
it is well known that the assembly of vortices occurring in the
XY-model and the similar two-dimensional condensates
reveal the topological phase transition, known as the
Kosterlitz Thouless (KT) transition
\cite{Kost}.
This shows the occurrence of
dissolution of vortex-antivortex pairs at some finite temperature.
This implies also the vortex pair excitations as
the fluctuation out of a state with quasi long range order
in two-dimensional XY and superfluid models.

The purpose of this paper is to study the quantum statistical
mechanics of 2-dimensional vortex gas on the basis of the generalized
Hamiltonian dynamics that is developed in the previous two papers.
Specifically we are concerned with
the statistical properties of vortex-antivortex pairs.
We are first concerned with the quantization of a vortex pair
by using the Bohr-Sommerfeld quantization scheme.
By using the resultant quantized spectra for a vortex pair,
we next evaluate the statistical partition function for a
vortex pair in a dilute gas limit.  We get a remarkable
consequence that the partition function is given in terms
of the generalized Riemann zeta function.
As the characteristic property of the zeta function we see that
the topological phase transition of KT  type is naturally
explained by the pole structure of the zeta function.

\bigskip
\bigskip

\noindent
{\large\bf Basic Formulation}

We start with a concise review of the previous results
for the later discussion.  The basic standpoint is that
the center of vortices $ (x_i, y_i) \equiv {\vec X}_i $ are
regarded as canonical conjugate like variables.
The effective action for the center of vortices is given by
\begin{equation}
S_{\rm eff} = \int (\sum_i (A_i{dx_i \over dt}
              - B_i{dy_i \over dt}) - H_{\rm eff}) dt .
\label{aa}
\end{equation}
Here the first term is called the canonical term and the
coefficients $ A_i , B_i $ are given in terms of the
\lq\lq density function\rq\rq , say $ \sigma $,
 together with the function $ \alpha_i \equiv
\log\vert \vec x -{\vec X}_i\vert $,
\footnote{
The definition of the density function $ \sigma $
is different for two cases of bose and spin fluids;
see ref. [2] and [3].})
whereas the second term
represents the effective Hamiltonian for the motion
of vortex centers which has the well known log potential form:
\begin{equation}
H_{\rm eff} = -{1 \over 2}\rho_0\sum_{ij} \mu_i \mu_j \log |{\vec X}_i
              - {\vec X}_j| + C
\label{ab}
\end{equation}
and hence the equation of motion for the vortex centers becomes
\begin{equation}
\sum_j G_{ij} {\dot {\vec X_j}} = - {\partial H_{\rm eff} \over
           \partial {\vec X}_i}
\label{ac}
\end{equation}
where $G$ is the metric tensor and $\mu$ denotes the strength
(charge) of the vortex.  $C$ represents the chemical potential
or self-energy that is given by a sum of the self-energy of
isolated vortices.  The tensor $G$ coincides with the
symplectic form (two-form) and it is derived from the
effective action $\omega$
\begin{equation}
d\omega = \sum_{ij}G_{ij}d{\vec X}_i \wedge d{\vec X}_j \\
\label{ad}
\end{equation}
with
\begin{equation}
G=\left[\matrix{g^{11} & -g^{12} \cr
          g^{12} & -g^{22} \cr}\right]
\label{ae}
\end{equation}
where each submatrix $g$ is given by the density function etc.
through the coefficients $ A_i, B_i $ in (\ref{aa})
and the concrete form for these is given in term of the
overlap integrals between the gradient of $ \sigma $ as well as
the $ \alpha $ functions as has been done in the previous papers
for the bose fluid and Heisenberg spin model.
Thus the physical meaning of these terms suggests the effect of
finite size of vortices. We see that
\begin{equation}
g^{11}=g^{22}=0, \quad g^{12}= g^0 + \Delta g .
\label{af}
\end{equation}
$g^0$ represents the diagonal and $\Delta g$ the
off-diagonal term of the matrix $g^{12}$ respectively
with respect to the vortex indices. We can introduce the poisson
bracket using the tensor $G$:
\begin{equation}
 \{A,B\}_{\rm PB} = \sum_{ij}\big(
    {\partial A \over \partial {\vec X}_i}
    {\partial B \over \partial {\vec X}_j}
    - \{{A\rightarrow B}\}\big) .
\label{ag}
\end{equation}
As a special case we have
\begin{equation}
\{ {\vec X}_i, {\vec X}_j\} = (G^{-1})_{ij} .
\label{ah}
\end{equation}
In the more concrete form,
\begin{equation}
 \{x_i , y_j\}_{\rm PB} = (g^{12})^{-1} = (g^0)^{-1}_{ij} +
 \{(g^0)^{-2} \Delta g \}_{ij} + \cdots .
\label{ai}
\end{equation}
The second term of LHS can be understood to mean that $\Delta g$
causes the anomalous effect.
Note that the above formulation can be regarded as an
important generalization of the classical Kirchhoff equation
$ \dot x =\{ x, H \}_{\rm PB} $ and $ \dot y = - \{y, H \}_{\rm PB}$ ,
with the Hamiltonian has the same logarithmic form, whereas
$\{A,B \}_{\rm PB}$ denotes the usual Poisson bracket.
The quantization is carried out by simply replacing the
generalized PB by the commutator: $ \{x_i,y_j\} \rightarrow
[x_i,y_j] $. Furthermore to be mentioned is the following fact:
The equation of motion for the case that $ \Delta g $ vanishes
is rewritten as
\begin{equation}
(g_0)_{ii}\;(\vec k\times \dot {\vec X}_i) =
{\partial H_{\rm eff} \over \partial {\vec X}_i}.
\label{a1x}
\end{equation}
This equation suggests that the canonical equation of motion
(Kirchhoff equation) is alternatively regarded as the
balance equation of two type \lq\lq forces\rq\rq , the one is the
right hand side, that represents the force derived from the
\lq\lq potential function\rq\rq \ $H_{\rm eff}$ and the other is the
left hand side, that represents the velocity of the $i$-th
vortex in a literal sense times the vorticity vector
\lq\lq $ g_0 {\vec k}$\rq\rq, where $ \vec k $ means the unit vector
perpendicular to $ (x,y) $ plane. The latter corresponds to
the so-called Magnus force.
\cite{Vinen}.

\bigskip
\bigskip

\noindent
{\large\bf Quantized Spectra for a Vortex pair}

We shall now treat the system of two vortices
whose charges are opposite each other, that is, $ \mu_1 = -\mu_2
\equiv \mu $  and quantize it.  The most interesting
quantity is the quantized energy spectrum for the vortex pair,
which will be used to evaluate the statistical function of the
vortex gas. In order to carry out this, we first consider the
case that the non-canonical effect is omitted and its effect
will be taken into account as a perturbation. By having
this in mind, we shall introduce the \lq\lq center of mass\rq\rq \ and
\lq\lq relative\rq\rq \ coordinates as
\begin{equation}
 x_1 + x_2 \equiv 2Q,\qquad y_1 + y_2 \equiv 2P
\label{ba}
\end{equation}
and
\begin{equation}
x_1 - x_2 \equiv q , \qquad y_1 - y_2 \equiv p
\label{bb}
\end{equation}
where $(x_i, y_i)$ is the coordinate of the center of $i$-th vortex.
The effective Hamiltonian which is derivatived previously becomes
\begin{equation}
 H_{\rm eff} \equiv E = {1 \over 4} \mu^2 \rho_2 \log(q^2 + p^2) +C~,
\label{bc}
\end{equation}
where the Hamiltonian is changed to positive definite.
\footnote{
It should be noted that this effective action do not have
the cut-off parameter.}
Hence we obtain the equations of motion as follows:
\begin{equation}
  {dQ \over dt} =0, \qquad {dP \over dt}=0,
\label{bd}
\end{equation}
together with
\begin{equation}
{dq \over dt} =  {\rho_0 \mu^2 p \over 2 \; (q^2 + p^2)}, \qquad
{dp \over dt} = -{\rho_0 \mu^2 q \over 2 \; (q^2 + p^2)}.
\label{be}
\end{equation}
 From the first set of equations, it follows that
$Q$ and $P$ become constant of the motion,
which means that the center of vortices is at rest.
On the other hand,  from the second set of equation
it follows that the orbit in the (p,q) plane
forms a circle: $ p^2 + q^2 = {\rm constant} $.

We shall carry out by using Bohr-Sommerfeld quatization, namely,
\begin{equation}
\oint_C \omega_0 =(n+{1\over 2})h .
\label{bf}
\end{equation}
where $n$ = integer and the factor the factor
$ {1\over 2} $ is added, which corresponds to the zero point
energy.
\footnote{
The more exotic quantum condition may be obtained,
which results in the quantum number of quater integer,
see \cite{Chiao}, but such a possibility is not considered hereafter.}
Physically speaking, the above quantization rule suggests
the quantization of the angular momentum carried by a vortex pair.
In this way, if we note that
the unperturbed orbit $ C $ for the relative motion becomes
a circle of radius , say, $ R $, the quantization for
non-perturbed case is calculated as
\begin{equation}
{g_0 \over 2} \int\int dq \wedge dp =  (n+{1\over 2})h
\label{bg}
\end{equation}
which leads to
\begin{equation}
{g^0 \over 2} \pi R^2 = (n + {1\over 2})h .
\label{bh}
\end{equation}
Hence the radius $ R $ takes a quantized value;
$ R_n^2 ={2 h \over g^0 \pi}n $ with $ n $ integers.
In the above expression, $ g_0 $ takes
$ g_0 = m\rho_0\mu $ (bose fluids), $ g_0 = J\mu\hbar $ (spin fluid).
Next, if the off-diagonal term is taken account of
as perturbation,
\begin{equation}
R_{n'}^2 = {2 \over g^0 \pi}(nh - \Gamma (R_{n'})) .
\label{bk}
\end{equation}
Here the value of $ \Gamma $ can be written as the function of $ R $.
Qualitatively, this becomes small as the size of the vortex
small and it tends to a constant value as $ \rightarrow \infty $
(the explicit evaluation may be given in the separate paper)
\cite{Ono2}. In this way,
$R_{n'}$ should be determined from the self-consistent equation,
in other words, the quantum number $n$ changes to the
modified value $n'$, which may be written as
\begin{equation}
n \rightarrow n'= n - {\Gamma (n') \over h}.
\label{bi}
\end{equation}
If we assume that the effect of $\Gamma$ is sufficiently small and
hence put $\Gamma(R_{n'})=-{\tilde \Gamma}={\rm constant}$, then
we have the simpler form for the quantum condition:
\begin{equation}
R_{n'}^2 = {2 \over g^0 \pi}\big( (n + {1 \over 2})h
           + \tilde\Gamma \big)
\label{bl}
\end{equation}
which yields
\begin{equation}
 n'= n + {\tilde \Gamma \over h} .
\label{bm}
\end{equation}
If we substitute (\ref{bm}) into (\ref{bc}),
we get the energy spectra for the relative motion:
\begin{eqnarray}
     E_n & = &{ 1 \over 4} \mu^2 \rho_0 \log R^2_{n'} + C \nonumber \\
         & = & {1 \over 4}\mu^2 \rho_0
                \log({2 \over g^0 \pi}
                \big( (n + {1 \over 2})h + \tilde\Gamma \big)) + C.
\label{bo}
\end{eqnarray}
Summarizing the above result, the angular momentum quantum
number does not take integer values.

\bigskip
\bigskip

\noindent
{\large\bf Partition Function of Dilute Vortex Gas}

Now we investigate statistical mechanics of two-dimensional gas of
vortices with strength $\mu$ (charge) in the medium
of \lq\lq charge\rq\rq \ neutrality and each vortex interacts
through $\log$ potential.
The thermodynamical quantities are obtained through the partition
function $ Z ={\rm Tr} \; {\rm e}^{-\beta{H_{\rm eff}}} $, where
$ \beta \equiv 1/kT $ and $T$ is
the temperature and $k$ denotes the Boltzmann factor.
The trace means the sum over all possible states
the system is allowed to take.
Here we restrict ourselves to a special situation: namely,
the chemical potential $C$,
which is necessary to create a pair of vortex of
plus and minus charge,  is very large. In this extreme
situation, the vortex system can be treated as a dilute gas
consisting vortex pairs.
In this limit, it suffices to treat only one pair of vortices
to consider the partition function. This is easily evaluated
to be
\begin{eqnarray}
  Z &=& \sum_{n=0}^\infty \exp\left[ - \beta H_{\rm eff} \right] \nonumber \\
    & = & \sum_{n=0}^\infty \exp \left[ -\beta
         \left\{{1 \over 4} \mu^2 \rho_0 \log({2 \over g^0 \pi}
         \big( (n + {1 \over 2})h + \tilde\Gamma \big))
          +C \right\} \right]
\label{bp}
\end{eqnarray}
which leads to
\begin{equation}
Z = \left( {4 \hbar \over g^0} \right)^{-{\beta \mu^2 \rho_0 \over 4}}
         \zeta({\beta \mu^2 \rho_0 \over 4}, {1 \over 2} +
          {{\tilde \Gamma}\over h}) \exp [-\beta C]
\label{bq}
\end{equation}
and the free energy of the system is given by
\begin{equation}
F = -kT\log Z .
\label{bqx}
\end{equation}
Here $ \zeta(s,a) $ is defined by
\begin{equation}
\zeta(s, a) = \sum_{n=0}^\infty {1 \over (n+a)^s}
\label{br}
\end{equation}
which is just a generalized form of Riemann zeta function
(in mathematical literature, this is called \lq\lq Hurwitz zeta function").
\cite{zeta}
The conventional zeta function is obtained by setting
$ \tilde\Gamma = 0 $. Remarkable is that $ \zeta $ diverges at
the point at $s=1$;
namely, $ \zeta $ function has a simple pole at $ s=1 $.
This pole singularity causes the divergence of the free energy
at finite temperature; ${\beta \mu^2 \rho_0 \over 4}=1$, namely,
the critical temperature is given by $ T'_c = {\mu^2 \rho_0 \over 4k} $
This fact suggests that a phase transition occurs at this
temperature. We shall also examine the susceptibility due to
these pair, which is given by
\begin{equation}
\chi = {1 \over 2}\mu^2\beta <R^2>
\end{equation}
where $<R^2>$ represents the mean square separation of
the each dipole pair, which becomes
\begin{eqnarray}
<R^2> &=& {1 \over Z}{\rm Tr} \{ R^2 \exp[-\beta H_{\rm eff}] \} \nonumber \\
      &=& {\zeta ({\beta \mu^2 \rho_0 \over 4}-1,
          {1 \over 2} + {\tilde\Gamma \over h}) \over
\zeta ({\beta \mu^2 \rho_0 \over 4}, {1 \over 2} + {\tilde\Gamma \over h})}.
\label{bs}
\end{eqnarray}

The above expression shows up the following remarkable consequence;
the pole of (\ref{bs}), namely, ${\beta \mu^2 \rho_0 \over 4}-1=1$,
defines another transition temperature: If the temperature
approaches to $T_c$ from the below, the mean separation between
the pair diverges at this temperature.
We see that this position corresponds to nothing but
the temperature $T_c$ of the KT transition at which a dipole pair
is made to dissociate.  Note that this temperature
is just half of $ T'_c $ noted in the above.
The free energy or partition function itself
remains finite at this temperature, namely, this
feature is seen from the formula
\begin{equation}
  Z = \zeta(2, {1 \over 2} + {\tilde\Gamma \over h})
\end{equation}
which is finite. In the above argument we assume that
the position of pole and hence the transition temperature
is not affected by the presence of
shift of quantum number by an amount of $ \tilde\Gamma $,
which follows from the theorem concerning the Hurwitz zeta function.

\bigskip
\bigskip

\noindent
{\large\bf Summary}

To summarize the present result is that after having
calculated the energy spectra including the effect of
the non-canonical term coming from the specific nature of
quantum vortices, we have shown that the topological phase
transition of KT type is characterized by the pole structure of the
generalized Riemann zeta function.  Up to our knowledge,
this fact has not been known by anyone previously,
which suggests that characteristics of statistical thermodynamics
of vortex gas is made clear
by exploiting the specific properties of zeta function.
In this connection, we note the appearance of two different transition
temperatures; $ T_c $ and $ T'_c $. It seems puzzled to have two
different transition temperatures, but it should be
noted that the present theory is based a rather naive assumption,
that is, only one vortex pair is considered and no correlation,
which arises from the other vortex pairs, is not taken into account.
If we incorporate this correlation in a proper manner, we
may have a correct value of the temperature for the
topological phase transition.
As the other problems we must consider the case that
the noncanonical term has dependence of the quantum number $ n $
which is shown by the self-consistent nature of the Bohr-Sommerfeld
equation. The details of these discussion will be given
elsewhere.

\bigskip
One of the authors (H.K) would like to thank Dr. H.Yabu for his useful
comment on the asymptotic formula for the Riemann zeta function.
They are also grateful to Prof. Iida for his useful discussion.

\end{document}